\documentclass[aps,twocolumn,showpacs,amsmath,amssymb,pra,superscriptaddress,floatfix,longbibliography]{revtex4-1}

\usepackage{times}
\usepackage{t1enc}
\usepackage{subfigure}
\usepackage{graphicx}
\usepackage{epstopdf}
\usepackage{amsmath}
\usepackage{dcolumn}
\usepackage{xcolor} 
\usepackage{soul}
\usepackage{bbold}
\usepackage{braket}
\usepackage{hyperref}

\hypersetup{
    colorlinks=true,
    linkcolor=blue,
    filecolor=magenta,      
    urlcolor=blue,
    citecolor = blue,
    }

\begin{document}


\title{Collective emission and selective radiance in atomic clouds and arrays coupled to a microring resonator}

\author{Deepak A.~Suresh}
\affiliation{Department of Physics and Astronomy, Purdue University, West Lafayette,
Indiana 47907, USA}

\author{Xinchao Zhou}
\affiliation{Department of Physics and Astronomy, Purdue University, West Lafayette,
Indiana 47907, USA}

\author{Chen-Lung Hung}
\affiliation{Department of Physics and Astronomy, Purdue University, West Lafayette,
Indiana 47907, USA}
\affiliation{Purdue Quantum Science and Engineering Institute, Purdue
University, West Lafayette, Indiana 47907, USA}

\author{F.~Robicheaux}
\email{robichf@purdue.edu}
\affiliation{Department of Physics and Astronomy, Purdue University, West Lafayette,
Indiana 47907, USA}
\affiliation{Purdue Quantum Science and Engineering Institute, Purdue
University, West Lafayette, Indiana 47907, USA}




\date{\today}

\begin{abstract}
We theoretically investigate the collective dipole-dipole interactions in atoms coupled to a nanophotonic microring resonator. The atoms can interact with each other through light-induced dipole-dipole interactions mediated by free space and through the resonator whispering-gallery modes. The differing characteristics and mismatched wavenumbers of these modes give rise to complex dynamics and provide new opportunities for controlling light–matter interactions. We explore these phenomena in the context of an experimentally realized atom cloud and study the potential of the proposed sub-wavelength atom arrays.

\end{abstract}


\maketitle

\section{Introduction\label{sec:intro}}

The study of collective effects due to dipole-dipole interactions has been a major topic of interest since Dicke described the concept of superradiance in his seminal paper in 1954 \cite{dicke}. There has been abundant research following this on the different aspects of collective dipole interactions like superradiance, subradiance, and collective Lamb shifts \cite{bettles,re1971,fhm1973,gfp1976,s2009,mss2014,pbj2014,bbl2016,bzb2016,psr2017,jbs2018}. These effects have been utilized for applications in quantum control, photon storage, and quantum information \cite{bga2016,fjr2016,swl2017,mma2018}. 

Great progress has been achieved in engineering nanophotonic interfaces for versatile control over light-matter interaction.
The interaction between atoms has been experimentally implemented using nanofibers \cite{sbf2017,lr2017,pb2022,pl2022,lp2023,lt2024,bt2024,sj2023}, photonic crystals \cite{gh2015,hg2016}, and slot waveguides \cite{rg2018,as2024}. Since the atoms are usually trapped close to dielectric surfaces, the complexities of modeling the interactions have been explored in Refs. \cite{ama2017,jn2018,so2023,cs2023}.

Recently, neutral atoms have been laser-cooled and trapped adjacent to a nanophotonic microring resonator \cite{zh2024} to be used as a versatile light-matter interaction platform. The ring-resonator can act as a whispering gallery mode (WGM) cavity and facilitate chiral atom-light interactions.

This platform has several unique properties. The atoms can interact with each other through two different types of interaction, through free space and through the resonator, each with a different characteristic wavenumber. The interaction strength with the resonator can also be varied by adjusting the position of the atoms. 

One of the potentials of such an atom-nanophotonic interface is the concept of "selective radiance" as was described in Ref. \cite{ama2017}. These selectively radiant states can be simultaneously superradiant to a desired guided mode and subradiant to undesired or error-inducing modes.
Due to the nature of the coupling with the ring resonator, the atoms will superradiantly couple with the resonator, and the decay rate will scale with the atom number \cite{zh2024}.
If the decay into free space can be reduced due to the collective dipole-dipole interactions, the interface becomes more robust against error from spontaneous emission into the vacuum. Such high-wavenumber subradiant states are naturally supported and accessed via the high refractive index ring resonator.

This light-matter platform could also have potential applications involving quantum memories \cite{ga2007}, quantum simulation \cite{gm2011,rp2014,pr2015,rv2016}, quantum networks \cite{sr2012}, and applications in quantum computing \cite{pk2016,gp2015,gp2017}. It holds the potential for investigating photon interactions \cite{rw2017,zm2019,sp2023}, non-Markovian effects \cite{z2023}, and topological models \cite{zm2020}.

In this paper, we theoretically and computationally model the atom and ring resonator system described in Ref. \cite{ExpPaper} to understand the physics underlying the interactions and explore further possibilities. 
We give particular emphasis on how the collective dipole-dipole interactions mediated by free space modes can affect the interaction with the microring resonator. 
We calculate and compare the decay dynamics of the photons emitted into free space and into the microring resonator.

We describe the system and the numerical methods used in Sec. \ref{sec:system} and Sec. \ref{sec:methods}, respectively. We simulate the current experiment, which involves a trapped atomic cloud above the ring resonator in Sec. \ref{sec:Cloud}. 
In Sec. \ref{sec:Arrays}, we explore the potential of an atom array near the ring resonator and consider the effect of disorder that might arise in experiments. 
This work serves as a companion paper to Ref. \cite{ExpPaper}, which describes the experimental aspects and discusses the measured photon decay rates.

\section{The System\label{sec:system}}

\begin{figure}
    \centering
    \includegraphics[width=1.0\columnwidth]{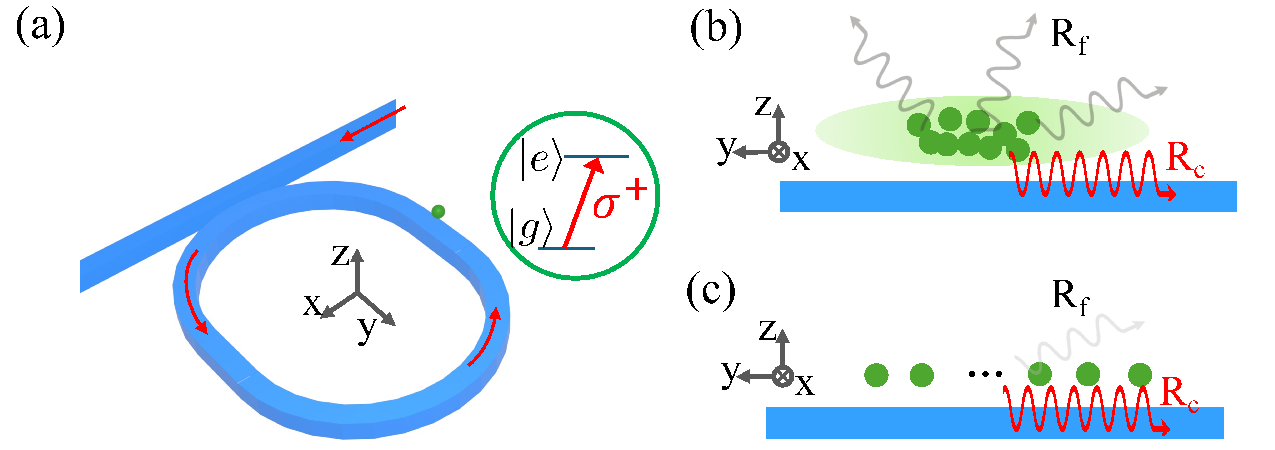}
    \caption{Schematic of the experimental system.
    (a) Cold atoms are trapped above the microring resonator, interacting with a WGM. The WGM is circularly polarized to drive the $\sigma^{+}$ cycling transition $\ket{g}\rightarrow\ket{e}$. The microring resonator has a total circumference of around $120 \mathrm{\mu m}$.
    (b,c) Interacting atomic ensemble (with $2 \mathrm{\mu m}$ r.m.s. along the y-direction) and organized atom array collectively emit photons into the WGM and the free space modes with photon emission rates $R_c$ and $R_f$, respectively. 
    }
    \label{fig:MRR}
\end{figure}

The experiment consists of a microring resonator coupled to a bus waveguide, as represented in Fig. \ref{fig:MRR}. 
The external coupling rate of the resonator to the bus waveguide ($\kappa_e$) is comparable to the intrinsic loss rate of the resonator ($\kappa_i$), which makes the resonator close to the critical coupling condition \cite{zh2023}. 
The microring resonator acts as a whispering gallery mode cavity that can sustain light circulating in the two directions. 

The experiment uses the fundamental Transverse Magnetic-like (TM) WGM. The electric field of the WGM is circularly polarized above the waveguide, and the polarization is locked to the direction of circulation in the resonator.
The atoms are polarized to a stretched state and coupled to the resonator mode via a cycling transition. As such, the atoms are constrained to emit into the same mode with which it is excited.
This configuration results in the atom-light interaction through the waveguide being chiral.
We only consider interaction with a fixed direction of circulation, as in Fig. \ref{fig:MRR}(a).

Cold atoms are cooled and trapped above the microring resonator with a peak density of around $10^{13}\,\mathrm{cm^{-3}}$ in an almost cigar-shaped configuration. The number of atoms in the cloud can be varied from a few to up to 60 atoms. 
These atoms are coupled to the evanescent field of the circulating WGM in the resonator. Since the evanescent field decreases exponentially above the ring, each atom has a different interaction strength with the WGM. This interaction strength can also be controlled by moving the atom cloud closer or farther away from the resonator.

To model this system, we use the master equation to evolve the density matrix of the system given by 
\begin{equation}
    \frac{d\hat{\rho}}{dt} = -\frac{i}{\hbar}[\hat{H}_{{eff}}, \hat{\rho}] + \mathcal{L} (\hat{\rho}),
    \label{eqn:Master}
\end{equation}
where $\hbar$ is the reduced Planck constant, $\hat{\rho}$ is the density matrix of the system with the Hilbert space only comprising the atomic excitations. $\hat{H}_{{eff}}$ is the effective Hamiltonian and $\mathcal{L}(\hat{\rho})$ is the Lindblad super-operator which describes the lossy interaction. These have separate contributions from the two types of interactions and are given by,
\begin{equation}
    \hat{H}_{{eff}} = \hat{H}_{dd}^f + \hat{H}_{dd}^c + \hat{H}_L^c
    \label{eqn:Heff}
\end{equation}
\begin{equation}
    \mathcal{L}(\hat{\rho}) = \mathcal{L}_f(\hat{\rho}) + \mathcal{L}_c(\hat{\rho}).
    \label{eqn:Leff}
\end{equation}

The terms $\hat{H}_{dd}^f$ and $\mathcal{L}_f(\hat{\rho})$ describe the atom–atom interactions mediated by the free-space modes. $\hat{H}_{dd}^c$, $\hat{H}_L^c$, and $\mathcal{L}_c(\hat{\rho})$ represent the effective interactions mediated by the microring cavity, derived via adiabatic elimination of the cavity photon degree of freedom. These mechanisms are described in detail in the following Sections \ref{sec:dipoleInteraction} and \ref{sec:cavityInteraction}.

Since we are restricted to the low-intensity limit, there can be only one excitation in the system at any time. We consider $N$ atoms in the system, and the Hilbert space consists of the collective ground state and the $N$ states in the single excitation manifold. The raising and lowering operators of the $j^\mathrm{th}$ atom are represented by $\hat{\sigma}^+_j$ and $\hat{\sigma}^-_j$ respectively.
The state in which all atoms are in the ground state is represented by $|g\rangle$, and the states where only the $j^\mathrm{th}$ atom is excited will be represented by $|e_j\rangle = \hat{\sigma}^+_j|g\rangle$. The single-atom decay rate of the atoms into free space will be denoted by $\Gamma_0$ and will serve to set the timescale of the system.

We first describe the collective dipole-dipole interactions mediated by the non-guided modes in free space, followed by the effect of coupling the atomic ensemble to the cavity.

\subsection{Free space collective dipole-dipole interaction\label{sec:dipoleInteraction}}

The separations between adjacent atoms are small and comparable to the wavelength of light, leading to the emergence of collective dipole-dipole interactions through free space. This can be modeled using the free space dyadic Green's function $\mathbf{G}(\mathbf{r}_i,\mathbf{r}_j,\omega_0)$, where $\omega_0$ is the resonant light frequency and $\mathbf{r}_i$, $\mathbf{r}_j$ are the position vectors of atom $i$ and $j$.

The dispersive and dissipative interaction strengths $J_{ij}$ and $\Gamma_{ij}$ are given by the Green's function as
\begin{align}
    J_{ij}&=-\frac{\mu_0\omega_0^2}{\hbar}\vec{d}^{*}\cdot \mathrm{Re} \{\mathbf{G}(\mathbf{r}_i,\mathbf{r}_j,\omega_0) \}\cdot \vec{d}\\
    \Gamma_{ij}&=\frac{2\mu_0\omega_0^2}{\hbar} \vec{d}^{*}\cdot \mathrm{Im} \{\mathbf{G}(\mathbf{r}_i,\mathbf{r}_j,\omega_0)\}\cdot \vec{d},
\end{align}
where $\mu_0$ is the vacuum permeability and $\vec{d}$ is the dipole operator of the atoms.
The dispersive part of the Green's function induces a resonant dipole-dipole exchange Hamiltonian,
\begin{equation}
    \hat{H}_{dd}^f = \hbar \sum_{i \neq j} J_{ij} \hat{\sigma}_i^+ \hat{\sigma}_j^-,
    \label{eqn:Hdd}
\end{equation}
and the dissipative part of the free space interaction is described by the Lindblad super-operator,
\begin{equation}
    \mathcal{L}_{f}(\hat{\rho}) = \sum_{i,j}\frac{\Gamma_{ij}}{2}\big[
    2\hat{\sigma}_j^- \hat{\rho} \hat{\sigma}_i^+  
    - \hat{\sigma}_i^+ \hat{\sigma}_j^- \hat{\rho}
    - \hat{\rho}\hat{\sigma}_i^+ \hat{\sigma}_j^-
    \big].
    \label{eqn:lindbladf}
\end{equation}
The interaction matrix associated with the free space Green's function $G_{ij}^f$ can be written as
\begin{equation}
    G^{f}_{ij} = J_{ij} - i\frac{\Gamma_{ij}}{2}.
    \label{eqn:GreenF}
\end{equation}
The effects of the atoms being in the vicinity of the dielectrics are negligible in this case due to the distance from the surface ($z\gtrsim \lambda_0/2$, where $\lambda_0$ is the resonant wavelength). More details can be found in the Supplementary Material of Ref. \cite{ExpPaper}. Therefore, it would suffice to take the simple vacuum Green's function to make effective calculations. The interaction matrix $G_{ij}^f$ will become
\begin{equation}
\begin{split}
     G^{f}_{ij} = -i\frac{\Gamma_0}{2}\bigg[h_0^{(1)}&( k_0r_{ij})
     +\frac{3(\hat{r_{ij}}\cdot\hat{d})(\hat{r_{ij}}\cdot\hat{d}^*) - 1}{2} h_2^{(1)}(k_0r_{ij})
     \bigg],
     \label{eqn:greens}
\end{split}
\end{equation}
where  $\hat{r}_{ij} = \mathbf{r}_{ij}/r_{ij}$ is the unit vector along $\mathbf{r}_{ij}  = \mathbf{r}_i - \mathbf{r}_j$, $r_{ij} = |\mathbf{r}_{ij} |$ is the norm, and $h_l^{(1)}$ are the outgoing spherical Hankel function of angular momentum $l$. $h_0^{(1)}(x)=e^{ix}/(ix)$ and $h_2^{(1)}(x) = e^{ix}(-3i/x^3 - 3/x^2 + i/x)$. 
The coefficient in front of $h_2^{(1)}$ depends on the orientation of the dipoles and the driven transitions, which will be $(1-3\cos^2{\theta})/4$ for $\sigma^+$ circularly polarized light quantized in the x-direction.

\subsection{Interaction with the cavity\label{sec:cavityInteraction}}

When the atomic ensemble is brought to the vicinity of the microring resonator, the atoms can couple with the evanescent field of the cavity and interact.
We consider a single mode cavity supporting the $\sigma^+$ circularly polarized mode.
This cavity interaction can be modeled as a harmonic oscillator with raising (lowering) operators, $\hat{a}^{\dagger}(\hat{a})$.
The Hamiltonian of the atom-cavity interaction can be represented as
\begin{equation}
\begin{split}
    \hat{H}_I = &\sum_j g_j  (e^{-i \phi_j}\hat{a}\hat{\sigma}^+_j + e^{i \phi_j}\hat{a}^{\dagger}\hat{\sigma}^-_j)\\ &- \sum_j \Delta_A \hat{\sigma}_j^+\hat{\sigma}_j^-
    + \eta (\hat{a} +\hat{a}^{\dagger}) -\Delta_C \hat{a}^{\dagger}\hat{a},
    \label{eqn:H_I}
\end{split}
\end{equation}
where $g_j$ depends on the evanescent electric field at the position of atom $j$,  $\Delta_A = \omega - \omega_A$ and $\Delta_C = \omega - \omega_C$ are the detuning of the atomic resonance and the cavity resonance from the input light frequency, respectively.
$\eta$ is the classical driving rate due to the external driving through the bus waveguide.
$\phi_j = \mathbf{k}\cdot \mathbf{r}_j $, where $\mathbf{k}$ is the wavevector of the ring resonator mode at position $\mathbf{r}_j$. Here, $\vert \mathbf{k}\vert = k_\mathrm{wg}$ is the wavenumber of the waveguide mode; $k_\mathrm{wg}>k_0$, the wavenumber associated with the resonant wavelength of light. 

In the experiment, single-atom cooperativity plays a more relevant role in describing the strength of the atom-cavity interaction. The single atom cooperativity of atom $j$ is given by,
\begin{equation}
    C^{j} = \frac{4 g_j^2}{(\kappa_i+\kappa_e)\Gamma_0},
    \label{eqn:EvanescentCoupling}
\end{equation}
which depends on $g_j$ and in turn depends on the position of each atom. It decays exponentially as the distance between the atom and the resonator surface increases.
Given an atomic density distribution, the average single-atom cooperativity will be denoted as $C_1$. 
In general, the total cooperativity of the waveguide interaction scales as $N C_1$.

The microring resonator is in the bad-cavity limit, where the cavity dynamics happen at a much faster rate than the dynamics of the atomic internal states ($\kappa_e,\kappa_i \approx 100 \Gamma_0$). The cavity field can be adiabatically eliminated to more intuitively capture the dynamics of the internal excitations of the atoms. This will also drastically decrease the computational requirements.
By enforcing that the cavity reaches a steady state at every time step of the internal dynamic evolution, we can trace out the cavity modes and obtain a simplified model to study the interactions between the atoms. Further details on the adiabatic eliminations are explained in Appendix \ref{app:adiabaticElim}.

After eliminating the cavity, the interaction of the atoms through the cavity will have a form similar to the collective dipole interactions described in Sec. \ref{sec:dipoleInteraction}.
There will be a dipole-dipole exchange Hamiltonian and a loss-inducing Lindblad operator 
\begin{equation}
    \hat{H}_{dd}^c = \hbar \sum_{i, j}
    -\mathrm{Re}\bigg\{\frac{g_ig_j}{\tilde{\kappa}} \bigg\} e^{-i(\phi_i-\phi_j)}
    \hat{\sigma}_i^+ \hat{\sigma}_j^-
    \label{eqn:Hddc}
\end{equation}
\begin{equation}
\begin{split}
    \mathcal{L}_{c}(\hat{\rho}) = \sum_{i,j}
    &-\mathrm{Im}\bigg\{\frac{g_ig_j}{\tilde{\kappa}} \bigg\} e^{-i(\phi_i-\phi_j)} \times \\
    &\big[
    2\hat{\sigma}_j^- \hat{\rho} \hat{\sigma}_i^+  
    - \hat{\sigma}_i^+ \hat{\sigma}_j^- \hat{\rho}
    - \hat{\rho}\hat{\sigma}_i^+ \hat{\sigma}_j^-
    \big],
    \label{eqn:lindbladc}
\end{split}
\end{equation}
where $\tilde{\kappa}=\Delta_C+i(\kappa_i+\kappa_e)/2$. Similar to the free space Green's function Eq. \eqref{eqn:greens}, we define the interaction matrix through the cavity as
\begin{equation}
\begin{split}
     G^{c}_{ij} = \frac{g_i g_j}{\tilde{\kappa}} e^{-i(\phi_i-\phi_j)}.
     \label{eqn:greensC}
\end{split}
\end{equation}
Excluding the phase dependence $\phi_{i(j)}$ in the ring cavity, the Hamiltonian in Eq. \eqref{eqn:Hddc} and the dissipation terms in Eq. \eqref{eqn:lindbladc} are dictated by the real and imaginary parts of the interaction matrix in Eq. \eqref{eqn:greensC}, following the same form as Eqs. \eqref{eqn:Hdd}–\eqref{eqn:GreenF}.

The driving of the system through the bus waveguide will appear in a form similar to a rotating wave laser Hamiltonian
\begin{equation}
    \hat{H}_L^c =  \hbar \sum_j \left[ -\Delta_A\hat{\sigma}_j^+\hat{\sigma}_j^- + \frac{\Omega_j}{2}\hat{\sigma}_j^+ + \frac{\Omega^*_j}{2}\hat{\sigma}_j^- \right],
    \label{eqn:HL}
\end{equation}
where $\Omega_j$ is the coupling with which each atom is excited due to the evanescent mode of the microring. It can be connected with the classical driving rate $\eta$ as 
\begin{equation}
    \Omega_j =  -\frac{g_j e^{-i\phi_j}}{\tilde{\kappa} }\eta.
\end{equation}

\subsubsection{Evanescent coupling\label{sec:EvanescentCoupling}}

The atoms couple with the resonator mode through an evanescent coupling. The coupling decreases exponentially with distance from the resonator. 
The atoms' average position along the z-axis depends on the trapping potential, allowing control of the average single atom cooperativity $C_1$. This provides an opportunity to study the behavior of collective photon emissions as a function of the atom-cavity interaction strength.

The experiment controls the average position of the atoms to be $\sim400 \mathrm{nm}$ away from the surface, which corresponds to $C_1 \lesssim 0.05$. 
In the proposed plan to implement an atom array, the trapping methodology can afford the atoms to be much closer to the resonator waveguide, leading to much higher average cooperativities ($C_1 \sim 10$).

Although the position-dependent coupling can provide an opportunity for controlling the atom-WGM photon interaction strength, it can also cause issues when it comes to the randomized position of atoms in the cloud. The spread of the atoms in the z-direction due to the trapping potential will cause a stochastic non-uniformity in the single-atom cooperativity of the atoms. This causes a considerable amount of broadening and decoherence in the system.

\subsection{Comparison of the two interactions}

The two types of interactions discussed so far have distinct characteristics that make the system dynamics more interesting and complicated. The eigenmodes and eigenvalues of the corresponding Green's function matrices can provide an insight into the nature of the interactions. The real part describes the dispersive behavior, and the imaginary part describes the dissipative behavior.

The vacuum mode-mediated interaction between two atoms is symmetric. As a result, the Green's function of the free space interaction, $\mathbf{G}_f$, is a complex symmetric matrix ($\mathbf{G}_f^{\mathrm{T}}=\mathbf{G}_f$). It typically has non-zero real and imaginary eigenvalues and is hence both dissipative and dispersive.
However, the interaction through free-space has a characteristic wavenumber of $k_0=2\pi/\lambda_0$. This means that excitations which match this wavenumber can couple well with the free space radiation modes. Particularly in ordered arrays, excitations with wavenumber $k>k_0$ are beyond the light-cone ($ck>\omega_0$) and have suppressed dissipative emission into the vacuum light field.

The resonator-mediated interaction between two atoms is not symmetric, due to the fact that cavity photons circulate around the resonator like running waves.
At resonance ($\Delta_{C}=0$), the atom-cavity interaction matrix $\mathbf{G}_c$ is a skew-Hermitian matrix ($\mathbf{G}_c^{\dagger}=-\mathbf{G}_c$), with purely imaginary eigenvalues, making this interaction purely dissipative.
There is a single superradiant mode and $N-1$ completely dark subradiant modes. Exciting the atoms through the resonator will always couple to the superradiant mode with decay rate $N C_1\Gamma_0$ \cite{zh2024}. 
The characteristic wavenumber of the atom-cavity interaction will be set by the effective refractive index ($n_{\mathrm{eff}} > 1$) of the resonator waveguide, $k_{\mathrm{wg}} = n_{\mathrm{eff}} k_0$. This mismatch in wavenumber with the free space interaction causes the superradiant state of the atom-cavity interaction to be subradiant to free space in ordered systems.

\section{Model for emission dynamics\label{sec:methods}}

To understand the dynamics of the system, the density matrix can be time-evolved using the master equation shown in Eq. \eqref{eqn:Master} with  Runge-Kutta methods. 
The full cavity model and the adiabatically eliminated Lindblad model have both been numerically compared and verified in the single excitation limit. The limitation of adiabatic elimination is that the cavity cannot be in the strong-coupling regime. 

An important approximation is that the atoms are stationary within the time scale of the internal excitation decay. This is reasonable since the atoms in the cloud have a mean velocity of 3 cm/s, which corresponds to a couple of nanometers of movement in one excitation lifetime. Hence, to capture the dynamics of the ensemble, we can use a Monte Carlo method to take the average of many different possible positions for the atoms. In each iteration, the positions of the atoms are randomly assigned according to the configuration of the atom cloud. This gives rise to a distribution of measured values, as depicted in Fig. \ref{fig:Cl_Dec_Hist} (c,d). In all other parts of the manuscript, we focus on the ensemble-averaged values of the relevant observables.

Time evolving the master equation in this particular scenario can lead to a few complications in the calculations. Since the atom cloud is considerably dense, the atoms in the random positions could end up being relatively close to another atom. This will cause the free space Green's function to increase significantly, beyond the energy scale of the calculation, requiring prohibitively small time-steps. Additionally, since these systems tend to have some states with subradiant character, the time to reach the steady state is much longer than the timescale of typical excitation decay. These issues make it challenging to reach convergence when using more than a few atoms. Hence, we utilize the weak field limit and can derive a procedure to directly calculate the final state expectation values using eigenmode decomposition. 

The equation of motion of the expectation values at low intensities can be written as
\begin{equation}
\begin{split}
    \frac{d\langle\hat{\sigma}^-_{i}\rangle}{dt} = \quad i \Delta_A\langle\hat{\sigma}^- _{i}\rangle + i\Omega_i
    -i \sum_{j}(G_{ij}^c + G_{ij}^f)\langle\hat{\sigma}^-_{j}\rangle ,
    \label{eqn:MFTExpEqn}
\end{split}
\end{equation}
By rewriting the $\langle\hat{\sigma}^-_j\rangle$ as a vector,
\begin{equation}
    \vec{\sigma} = \{\langle\hat{\sigma}^-_1\rangle, ..., \langle\hat{\sigma}^-_N\rangle\},
\end{equation}
Eq. \eqref{eqn:MFTExpEqn} can be simplified as
\begin{equation}
    \dot{\vec{\sigma}} = i\mathbf{M}\vec{\sigma} +  i\vec{\Omega} ,
    \label{eqn:MFTEqn}
\end{equation}
where $\vec{\Omega} = \{\Omega_1, ..., \Omega_N\}$, and $\mathbf{M} = \Delta_A \mathbb{1} - \mathbf{G}_c - \mathbf{G}_f$ is the combined coupling matrix of the two types of interactions.

This matrix $\mathbf{M}$ can be diagonalized with N eigenvectors $\vec{v}_{\alpha}$ of complex eigenvalues 
\begin{equation}
    \mathbf{M}\vec{v}_{\alpha} = \lambda_{\alpha}\vec{v}_{\alpha}, \quad
    \lambda_{\alpha}=J_{\alpha}-i\frac{\Gamma_{\alpha}}{2}.
    \label{eqn:Eigenvalue}
\end{equation}
Here, the $\vec{v}_{\alpha}$ will be the right eigenvectors of the complex matrix $\mathbf{M}$.
An atomic state can be decomposed using the eigenvectors as 
\begin{equation}
    \vec{\sigma}=\sum_{\alpha}w_{\alpha}\vec{v}_{\alpha},
    \label{eqn:sigmavec}
\end{equation} 
where $w_\alpha$ is the weight associated with eigenmode $\alpha$. The final state after the evolution can be directly solved by separately solving the time evolution of each eigenmode. 

\subsection{Steady state vs timed-Dicke state}

In the experiment in Ref. \cite{ExpPaper}, the system is excited using either a short duration or a long duration pulse, creating a single excitation in the system. 
The collective excited states associated with these situations are considerably different.

In the case of the short pulse, the pulse time (typically around a few nanoseconds) is much shorter than the lifetime of the atomic excitation $(30 \mathrm{ns})$.
Hence, the pulse can be modelled to be infinitesimally short compared to the timescale of atomic excitation dynamics. The atoms are excited into a state similar to a spin-wave excitation known as the timed-Dicke state (TDS), which is a collective state with a well-defined wave-vector 
\cite{TDS2009}. Due to the higher refractive index of the resonator waveguide, the excited TDS will have a wave-vector corresponding to that of the resonator mode $(\mathbf{k}_{\mathrm{wg}})$ denoted by,
\begin{equation}
    \vert TDS \rangle = \frac{1}{\sqrt{\mathrm{N}}} \sum_j c_j e^{i \mathbf{k}_{\mathrm{wg}}\cdot\mathbf{r}_j} \vert e_j \rangle,
\end{equation}
where the coefficient $c_j$ is dependent on the individual atom cooperativities $C^j$.
This will be a superradiant state of the atom-cavity interaction. For decay into free space, the TDS has a higher wavenumber $k_{\mathrm{wg}}$ than $k_0$, and is typically associated with subradiance in ordered arrays.
For the calculation, the final state can be directly initialized into a timed-Dicke state (TDS) with a small excitation population since the pulse time will be too short to cause mixing into any other states, $\vec{\sigma}_{\mathrm{TDS}}=\sum_{\alpha}w_{\alpha}^{\mathrm{TDS}}\vec{v}_{\alpha}\propto\vec{\Omega}$.

The system can also be driven into the steady state (SS) by using a significantly longer pulse duration ($\sim200\mathrm {ns}$ in the experiment). 
Since there is constant driving, the system is usually pumped into excited states exhibiting slower decay. 
The long excitation facilitates a mixing between different modes, resulting in a range of different wave-vectors, instead of a single dominant wave-vector in the case of the TDS.
The detuning also plays an important role in determining which modes are excited in this case. The final state  $\vec{\sigma}_{\mathrm{SS}}=\sum_{\alpha}w_{\alpha}^{\mathrm{SS}}\vec{v}_{\alpha}$ can be calculated by assuming an infinitely long excitation until reaching $\dot{\vec{\sigma}}=0$.

The method to calculate $w_{\alpha}^\mathrm{SS}$ and $w_{\alpha}^\mathrm{TDS}$ and the non-standard orthogonality relations used are described in Appendix \ref{app:ortho}.

Using these methods to directly calculate the final state of the system instead of time propagating the density matrix greatly decreases the computation time. The convergence of the results from the mean field, adiabatic elimination, and the full density matrix has been numerically verified in the weak field limit.

\subsection{Photon emission into the cavity and free space\label{sec:Rates}}
We study the evolution of the system right after the excitation pulse is switched off. The system will then evolve under the equation $\dot{\vec{\sigma}} = i \mathbf{M}\vec{\sigma}$. For an initial state $\vec{\sigma}_0=\sum_{\alpha}w_{\alpha}\vec{v}_{\alpha}$ at $t=0$, the evolution will be $\vec{\sigma}(t)=\sum_{\alpha}w_{\alpha}e^{i\lambda_{\alpha}t}\vec{v}_{\alpha}$. Atomic excitation will decay to the ground state and emit photons into different modes. 
The rate of photon emission can be calculated from the total de-excitation rate in the system.,
\begin{equation}
\begin{split}
    R(t) &=  - \langle \dot{\hat{e}} \rangle = -\frac{d(\vec{\sigma}^{\dagger}\vec{\sigma})}{dt} = \vec{\sigma}^\dagger(i\mathbf{M}^\dagger - i\mathbf{M})\vec{\sigma}\\
    &= 2i\vec{\sigma}^\dagger\mathbf{G}_c\vec{\sigma} 
    - 2i\vec{\sigma}^\dagger \mathrm{Im}\{\mathbf{G}_f \}\vec{\sigma},
\end{split}
\end{equation}
where $\langle \hat{e} \rangle$ is the total excitation probability in the system, and we have used $\mathbf{G}_c^\dagger = - \mathbf{G}_c$ and $\mathbf{G}_f^T =  \mathbf{G}_f$. 
The time dependence of $\langle \hat{e} \rangle$ and $\vec{\sigma}$ is not explicitly written for the sake of simplicity.
The rate of photons emitted into the cavity mode and the free space modes can be separately calculated as
\begin{equation}
    R_{c}(t) =  2i\vec{\sigma}^{\dagger}\mathbf{G}_{c}\vec{\sigma}
\end{equation}
\begin{equation}
    R_{f}(t) =  - 2\vec{\sigma}^{\dagger}\mathrm{Im}\{\mathbf{G}_{f}\}\vec{\sigma}.
\end{equation}
The experimentally measured transmission rate at the output of the bus waveguide is directly proportional to the photon emission rate $R_c$. 
Photon emission into free space is typically collected by optics spanning over a finite solid angle. If the angular pattern of photon emission is fixed in time, this measured photon rate is proportional to $R_f$.

Although the amplitude associated with each eigenstate decays exponentially in time, the emission rates contain the interference between all the eigenmodes, leading to more complicated time evolution dynamics.
Hence, the total excitation probability does not decay purely exponentially. Correspondingly, the photon emission rates, as well as the decay rates, become time-dependent.
In this paper, the focus will be on the instantaneous decay rate immediately after the drive is turned off in order to study the initial time behaviors ($t \ll 1/\Gamma_0$) of the system. The decay rates of the excitation into the free space and cavity modes have been defined as
\begin{equation}
    \gamma_f = \frac{R_f(0)}{\langle \hat{e} \rangle(0)}\quad
    \mathrm{and}\quad
    \gamma_c = \frac{R_c(0)}{\langle \hat{e} \rangle(0)}.
    \label{eqn:gammafc}
\end{equation}
The former is approximately independent of the atom-cavity interaction strength, i.e., the average single atom cooperativity $C_1$. This is because, at low intensities, the decay rate into free space is solely dependent on the relative phase and the relative amplitude of the excitation between the atoms. 
While shifting the mean atomic position in the z-axis and changing the atom-cavity interaction strength could change the absolute excitation probability, it does not change the relative amplitude and phase between the atoms.

\subsubsection{Decay rates of photon emission}

The tools available in the experimental setup limits the reliable measurement of the photon decay rate to the cavity ($\dot{R}_c/R_c$). The photon emission rate into free space was not measured \cite{zh2024}. In systems with exponential decay, this measurement is equivalent to the decay rate of the excitation in the system. 
However, this measurement and the decay rates defined in Eq. \eqref{eqn:gammafc}
are not equivalent for non-exponential decay dynamics. Reliably measuring the excitation decay rate, like $\gamma_c$ or $\gamma_f$, remains a challenge in most experiments. In this particular platform, where the interaction strength can be varied and non-exponential decay is present, the discrepancy can be larger. Hence, the experimental paper in Ref. \cite{ExpPaper} operationally defines the decay of the photon emission rate into the free space and cavity modes as
\begin{equation}
    \Gamma_f = \frac{-\dot{R}_f(0)}{R(0)} \quad
    \mathrm{and}
    \quad \Gamma_c = \frac{-\dot{R}_c(0)}{R(0)},
\end{equation}
where $R = R_f + R_c$ is the total photon emission rate. The measurement of the decay rate of $R_c$ can be related to these decay rates as
\begin{equation}
    \Gamma_{\mathrm{exp}} = -\frac{\dot{R}_c(0)}{R_c(0)} = \Gamma_c + \Gamma_f\theta,
    \label{eqn:GamExp}
\end{equation}
where $\theta = [\dot{R}_c(0)/R_c(0)]/[\dot{R}_f(0)/R_f(0)]$. 

It is important to note that the difference between $\gamma_c,\gamma_f$ and $\Gamma_c, \Gamma_f$ vanishes in the case of single exponential decay. This is generally true for photon emission from an ordered atom array, where the excitation is largely confined to a single eigenmode.

\section{Results}

We consider two main configurations: atoms in a randomly distributed cloud and an ordered array of atoms. We explore the potential of both these configurations in utilizing collective states to suppress emission into free space while promoting stronger coupling to the cavity mode. The primary focus will be on comparing the decay characteristics due to emission into free space and the cavity.

In the experiment, the most relevant observable is the time evolution of the photon emission rate into the cavity mode, which can be measured as a change in transmission rate from the output waveguide. 
These measurements can be utilized to verify the theoretical calculations. Further calculations can be performed to expand the scope and understanding of the experiment.

\subsection{Atom Cloud\label{sec:Cloud}}

\begin{figure}
    \centering
    \includegraphics[width=1.0\linewidth]{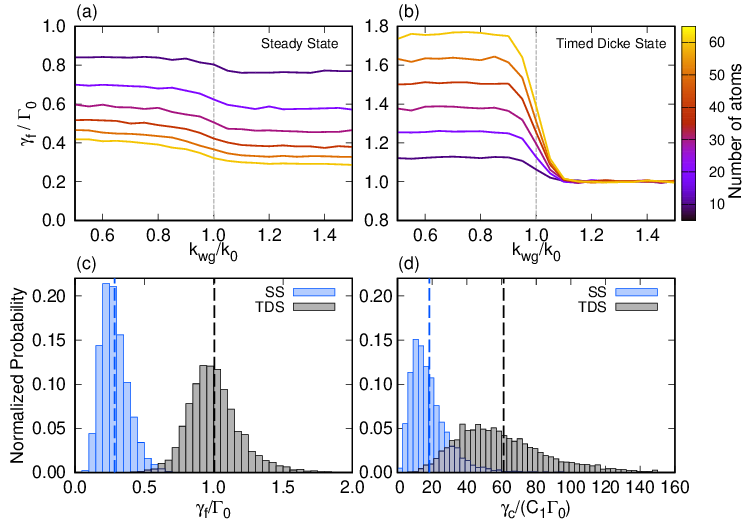}
    \caption{(a,b) Ensemble averaged decay rate into free space when an atom cloud is excited to (a) the steady state (SS) or (b) the timed-Dicke state (TDS) with a WGM of wavenumber $k_{\mathrm{wg}}$. The color scale depicts the number of atoms in the atom cloud. (c,d) Histograms of the distribution of the excitation decay rate into (c) free space and (d) into the cavity when taking individual random realizations of the atomic cloud with $N = 60$ atoms and $C_1 = 0.05$. The average value of the histogram has been plotted as vertical dashed lines of the corresponding color. The results are averages over 5000 random configurations.}
    \label{fig:Cl_Dec_Hist}
\end{figure}

\begin{figure}
    \centering
    \includegraphics[width=1.0\linewidth]{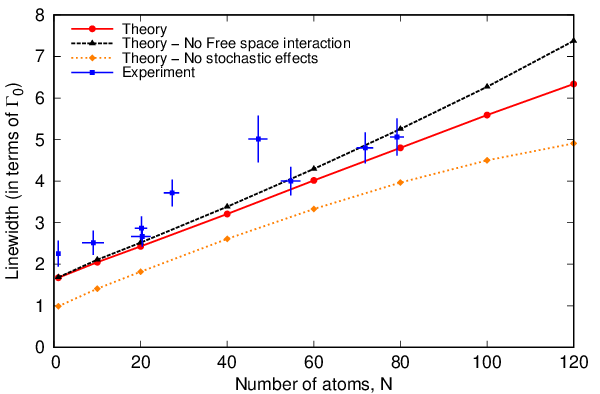}
    \caption{The measured and calculated linewidth of the steady state transmission spectrum as a function of the number of atoms for $C_1 = 0.05$. Blue squares with error bars denote experimental data. Red circles with lines show theoretical calculations. The black triangles and orange diamonds with dashed lines show the theoretical calculation of hypothetical situations where there is no free space collective dipole-dipole interaction or when there are no stochastic effects.}
    \label{fig:CloudLinewidth}
\end{figure}

\begin{figure}
    \centering
    \includegraphics[width=1.0\linewidth]{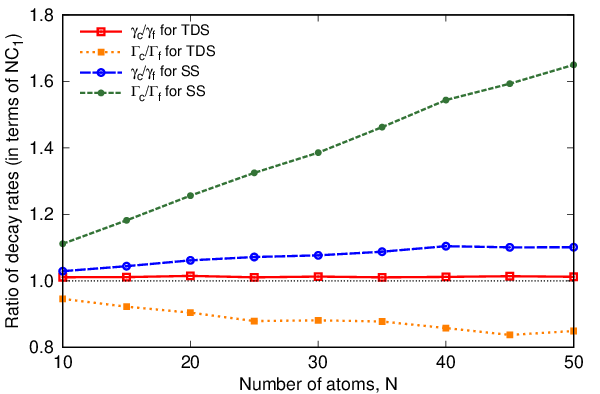}
    \caption{The ratio of decay rates for a cloud of atoms with uniform $C$ for the TDS and the SS. In both cases, the excitation decay ratio $\gamma_c/\gamma_f$ tends to be close to $NC_1$. The photon rate decay ratio $\Gamma_c/\Gamma_f$ increases or decreases depending on the type of excitation. The results are averaged over 2000 random configurations.}
    \label{fig:Cl_FOM}
\end{figure}

In the current state of the experiment, a cloud of atoms is trapped and cooled near the microring resonator in a trap with a cigar-shape with an r.m.s. of around $2 \mu \mathrm{m}$ along the waveguide. In the transverse direction, the cloud is tightly trapped to an r.m.s. of around $100 \mathrm{nm}$.
In the z-direction, the cloud has an approximate r.m.s. of $430 \mathrm{nm}$, but has an asymmetric density profile due to the shape of the trap, as detailed in Ref. \cite{zh2024}.

The system can be resonantly driven ($\omega = \omega_A$) with a large wavenumber ($k>k_0$). Unlike in free space, where systems are conventionally only driven by the free space modes with wavenumber $k_0$, the effective refractive index of the resonator waveguide determines the wavenumber of the drive. By increasing the excitation wavenumber, the rapid spatial variation in phase allows the system to access different collective excited states with varied decay rates.

Figure \ref{fig:Cl_Dec_Hist} (a) and (b) explore how driving an atomic cloud with a different wavenumber can affect the free space decay rate. 
The system is being excited using either short or long pulses, which can exhibit contrasting subradiant (SS) or superradiant (TDS) emission dynamics.

When driven using a short pulse, the system is excited into the TDS with the wavenumber $k \sim 1.7 k_0$, where $n_{\mathrm{eff}} \approx 1.7$ is numerically obtained based on the waveguide geometry presented in Ref. \cite{zh2024,ExpPaper}.
Although this state will be subradiant to free space in ordered arrays, Ref. \cite{gh2022} has shown that for an atom cloud, the smallest decay rate for such high-k states is $\Gamma_0$. This can be seen in Fig. \ref{fig:Cl_Dec_Hist} (b), where the free space decay rate $\gamma_f$ for all atom numbers converge to $\Gamma_0$ as the wavenumber increases beyond $k_0$. This happens because the dephasing from atoms being very close to each other in a random cloud acts against the longer-range coherence that is built up to cause subradiance. 
The TDS will be perfectly superradiant to the cavity with an average decay rate of $NC_1\Gamma_0$ while having a decay rate of $\Gamma_0$ into free space. Therefore, the TDS will have a total decay rate of $(1+NC_1) \Gamma_0$.

When the atom cloud reaches the SS, the system displays slower photon emission dynamics. As shown in Fig. \ref{fig:Cl_Dec_Hist} (a), the free-space decay rate exhibits subradiance and the rate decreases with an increasing number of atoms. While subradiance can occur even when the system is excited with a free-space wavenumber $k_0$, the subradiant effect is enhanced when driven with higher wavenumbers $k>k_0$.

These effects can also be seen in Fig. \ref{fig:Cl_Dec_Hist} (c) and (d), which show the distribution of the decay rates into the free space and cavity modes without ensemble averaging. The free space decay rate ($\gamma_f$) is suppressed in the case of the SS, while clustered around $\Gamma_0$ for the TDS. For the decay rate into the cavity ($\gamma_c$), the TDS has a large spread of decay rates, but the average is close to $N C_1 \Gamma_0$, where $N$ is the atom number in the system. On the other hand, for the SS, $\gamma_c$ averages to around $20 C_1 \Gamma_0$, which is still superradiant but not as good as in the case with the TDS. This shows that the TDS is better for utilizing superradiant effects.

The large spread in $\gamma_c$ is primarily due to the stochastic non-uniformity of the single atom cooperativity $C$, as mentioned in Sec. \ref{sec:EvanescentCoupling}. Instead, in the hypothetical situation where the atoms have no z-spread and the $C$ is uniform for all the atoms, the distribution of $\gamma_c$ will be a delta function at $NC_1\Gamma_0$ for the TDS, and will have a considerably smaller spread for the SS. This direct connection between the broadening of the spectrum and the spread in z-position emphasizes the importance of tight confinement for systems that have evanescent coupling.

On the other hand, the steady state transmission spectrum through the output bus waveguide can be calculated and compared with experimental measurements. The linewidth can provide some indication of the decay properties of the system. Figure \ref{fig:CloudLinewidth} compares the experimentally measured and calculated linewidths of the spectrum, which are both determined using simple Lorentzian fits. The mean atom number $N$ in the experiment is extracted by fitting the steady-state transmission spectrum assuming a Gamma distribution of $NC$ with mean value $C_N$ and $N=C_N/C_1$ (see Ref. \cite{zh2024}).

Without the free space interaction, the system will be excited to only the superradiant mode of the atom-cavity interaction, and the linewidth will scale as $NC_1$. This is depicted by the black dashed line. In the presence of free space collective interaction, it facilitates the mixing between the superradiant and the completely dark subradiant states of the atom-cavity interaction. This pulls down the total linewidth as $N$ increases. This can be seen more clearly from the theoretical data points.

However, in practical implementations, the expected linewidth suppression is significantly weakened due to multiple stochastic effects. The primary contributor is the non-uniformity in single-atom cooperativity $C$ (see Sec. \ref{sec:EvanescentCoupling}), which broadens the linewidth. Additionally, the AC Stark shift from trapping fields varies with atomic position, introducing a slight spectral asymmetry and further broadening the linewidth. Moreover, the number of trapped atoms follows a Poissonian distribution, which introduces further broadening, especially for small $N$. The orange diamonds in Fig. \ref{fig:CloudLinewidth} depict the hypothetical, where these stochastic effects are absent. This linewidth shows a greater degree of suppression for larger $N$ and approaches $\Gamma_0$ when $N \rightarrow 1$.

Another aspect of the spectrum is the average line-shift. The atom-cavity Green's function is purely dissipative and will not contribute to line-shifts. On the other hand, the interaction through free space will contribute to a shift depending on the density of the atoms. Therefore, a shift in the spectrum due to the free-space interaction can be expected. Contrary to expectations, the shifts observed experimentally were an order of magnitude higher than the calculations. Several effects, like the AC Stark shift, vicinity to a dielectric surface, and motion of the atoms were explored but were unable to explain the discrepancy. 
Much larger shifts are expected in the case of atom arrays, and clearer answers are expected once they are implemented in the next stage of the experiment. 

The ratio of decay rates into the desired versus the undesired modes describes the effectiveness in mitigating photon emission into the undesired channels. 
Figure \ref{fig:Cl_FOM} shows the ratio of decay rates for the excitation ($\gamma_c/\gamma_f$) and the photon decay rate ($\Gamma_c/\Gamma_f$). 
In this plot, we assume a uniform cooperativity $C$ to simplify numerical convergence. Incorporating non-uniformity in $C$ does not lead to qualitative differences.

In the case of the TDS, $\gamma_c/\gamma_f \sim NC_1\Gamma_0$, showing the same scaling when the atoms are far apart and collective dipole interactions through free space are absent. This behavior is already seen in Fig. \ref{fig:Cl_Dec_Hist}(c,d), where the average $\gamma_c/\Gamma_0 \approx NC_1$ and $\gamma_f \approx \Gamma_0$.
Surprisingly, in the case of the SS, although $\gamma_f$ is suppressed to be subradiant, $\gamma_c$ is also reduced to still give a scaling of $NC_1$ for the excitation decay ratio.

The photon rate decay ratio $\Gamma_c/\Gamma_f$ can exceed or fall below the expected $NC_1$ scaling depending on the excitation conditions. This happens because, although $\Gamma_c$ scales as $NC_1$ in both cases, $\Gamma_f$ increases or decreases with $N$ depending on the type of excitation. This can be seen in Fig. 4(c) in \cite{ExpPaper} where $\Gamma_f$ increases beyond $\Gamma_0$ with $N$ for the TDS and decreases below $\Gamma_0$ for the SS.

In the associated experimental paper \cite{ExpPaper}, the ratio $\theta$ in Eq. \eqref{eqn:GamExp} serves as an important factor to correlate experimentally measured decay rates to the actual decay rates of the system. To get an intuition behind this $\theta$, we can express it as $\theta = (\Gamma_c/\Gamma_f)/(\gamma_c/\gamma_f)$. 
In the case of the TDS, $\Gamma_c/\Gamma_f < NC_1$ resulting in the $\theta < 1$, while for the SS, $\Gamma_c/\Gamma_f > NC_1$ resulting in the $\theta > 1$. Further analysis on the implications of $\theta$ is explored in the Supplementary Material of Ref. \cite{ExpPaper}.

\subsection{Atom Arrays\label{sec:Arrays}}
\begin{figure}
    \centering
    \includegraphics[width=1.0\linewidth]{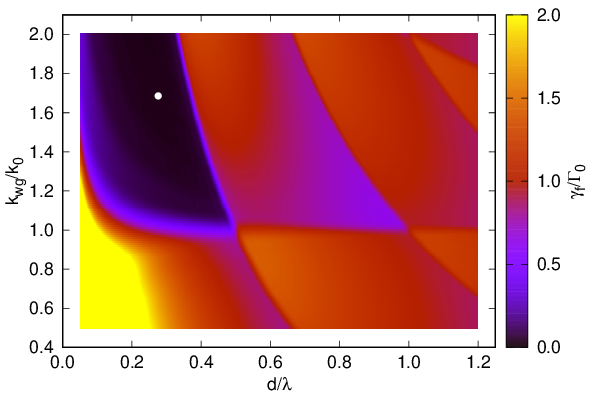}
    \caption{The decay rate into free space when a perfect atom array with 20 atoms is excited to the TDS through the resonator, as the separation $d$ and the WGM wavenumber $k_{\mathrm{wg}}$ are varied. The white point denotes the proposed parameters of the experiment, giving $\gamma_f = 0.035 \Gamma_0$.}
    \label{fig:FigAr_d_Neff}
\end{figure}

Although a disordered atomic ensemble can exhibit subradiance to free space under steady-state excitation, the effect is limited. In contrast, ordered arrays can mitigate randomness and promote the coherence necessary for pronounced subradiance. In this section, we explore the potential of such atom arrays in reaching subradiance in free space through the simple TDS excitation.

The next stage of the experiment involves the implementation of an array of atoms trapped above the ring resonator as depicted in Fig. \ref{fig:MRR} (c). According to \cite{cf2019}, the atoms can be trapped using two-color evanescent fields, resulting in the atoms being separated by about $0.3 \lambda_0$, where $\lambda_0$ is the resonant wavelength of the light in free space. 
This separation is approximately half the wavelength of the high-wavenumber guided mode in the ring resonator.

When the atoms are excited through the resonator, it can facilitate almost perfectly subradiant states into free space due to the alternating phases in adjacent atoms. 

The higher wavenumber of the waveguide excitation $k_{\mathrm{wg}}$ plays a major role in contributing to accessing highly subradiant states in the case of atom arrays. Experimentally, this is only dependent on the effective refractive index $n_{\mathrm{eff}} =k_{\mathrm{wg}}/k_0$ of the WGM. 
The decay rate into free space as a function of $k_{\mathrm{wg}}$ and the separation of the atom $d$ has been depicted in Fig. \ref{fig:FigAr_d_Neff} for a perfect atom array with no disorder. 
Although very good subradiance ($\gamma_f < 0.1\Gamma_0$) can only be achieved with $n_{\mathrm{eff}} > 1$ and separations $d < 0.5 \lambda_0$, there are still other regions where nominal subradiance ($\gamma_f < \Gamma_0$) can be achieved. As mentioned at the end of Sec. \ref{sec:Rates}, in the low-intensity limit, $\gamma_f$ is only dependent on the relative phase between the atomic excitation and is therefore independent of the average single atom cooperativity $C_1$.

The white dot in Fig. \ref{fig:FigAr_d_Neff} marks the proposed parameters of the upcoming experiment. For a perfect array with $N = 20$, separation of $d = 0.3 \lambda_0$, and effective refractive index $n_{\mathrm{eff}} = 1.69$,  the free space decay rate $\gamma_f = 0.035 \Gamma_0$. The atoms are assumed to be trapped around $330 \mathrm{nm}$ from the resonator to give an average $C_1 = 0.05$. If the atoms are trapped much closer, as is proposed to be possible by Ref. \cite{cf2019}, much larger cooperativities can be achieved. Modeling this would require modification to the free space Green's function due to the vicinity of the dielectric, as well as going beyond the adiabatic elimination approximation.

\subsubsection{Effects of Disorder\label{sec:ArDisorder}}

Although atom arrays show great promise in achieving subradiance in free space, inevitable defects due to experimental implementations can be limiting. We study how two common sources of imperfection can disturb the delicate coherences built in an ideal subradiant state.

The coupling of the cavity mode with the atoms has an exponential dependence on the z-position due to the nature of the evanescent coupling. Thus, the randomness in the z-position due to trapping will cause a stochastic non-uniformity in $C$ of each atom. This randomness in the single atom cooperativity can be treated as a source of decoherence.

\begin{figure}
    \centering
    \includegraphics[width=1.0\linewidth]{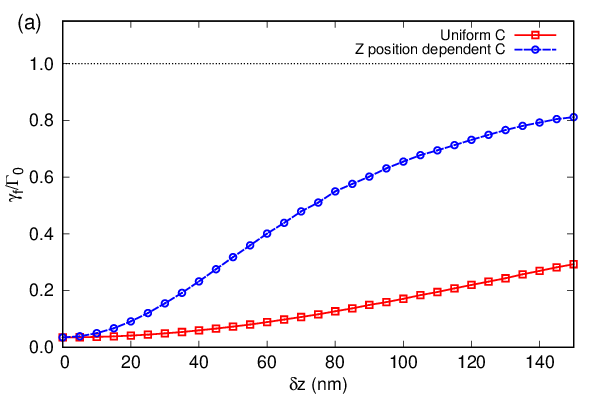}
    \includegraphics[width=1.0\linewidth]{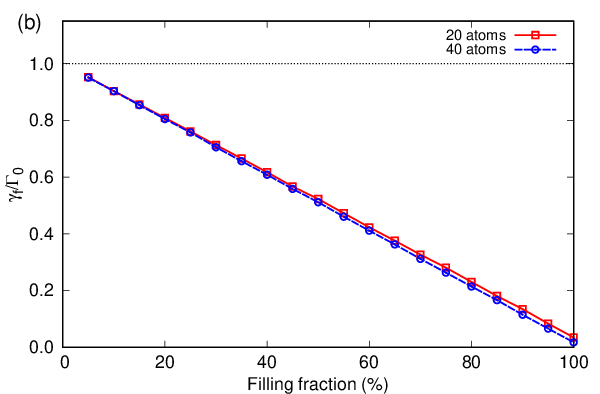}
    \caption{Effects of disorder in atom arrays with the TDS. The decay rate into free space has been plotted versus (a) the spread $\delta z$ along the z-axis and (b) the filling fraction.}
    \label{fig:Ar_Disorder}
\end{figure}

In Fig. \ref{fig:Ar_Disorder}(a), the decay rate into free space is plotted as a function of the r.m.s. size $\delta z$ of the atomic wavepacket along the z-axis. In this case, the spread in the other two directions is assumed to be negligible to isolate the impact of $\delta z$. The red squares denote the case when all the atoms have a uniform $C$, and the blue circles denote the situation where the $C$ is z-position dependent. Just the spread in position alone can cause decoherence and reduce the degree of subradiance, but the stochastic non-uniformity in $C$ causes a much larger effect.
At $\delta z = 50 \mathrm{nm}$ spread, the reduction in subradiance due to the non-uniformity in $C$ is around $8$ times larger than the reduction in the case with uniform $C$. This emphasizes the importance of confining the atoms tightly in the z-direction in the case of a system with evanescent coupling.

In Fig. \ref{fig:Ar_Disorder}(b), the effect of the filling fraction of the array is depicted. 
We grow the system size by probabilistically placing one atom in each added site based on the filling fraction until the system contains 20 or 40 atoms.
Although the number of array sites will vary with the filling fraction, this is adopted to maintain the scaling of the waveguide interaction with the number of atoms.
The trend is almost linear, connecting the points from $\Gamma_0$ asymptotically at $0\%$ filling fraction to the minimum possible decay rate at $100\%$ filling fraction. This emphasizes the importance of implementing arrays with a high filling fraction to achieve good subradiance in free space.
Even if the state is completely dark for a perfect array, a $50\%$ filling fraction can only allow a minimum of $\gamma_f = 0.5 \Gamma_0$.

\subsubsection{Circular Array}

\begin{figure}
    \centering
    \includegraphics[width=1.0\linewidth]{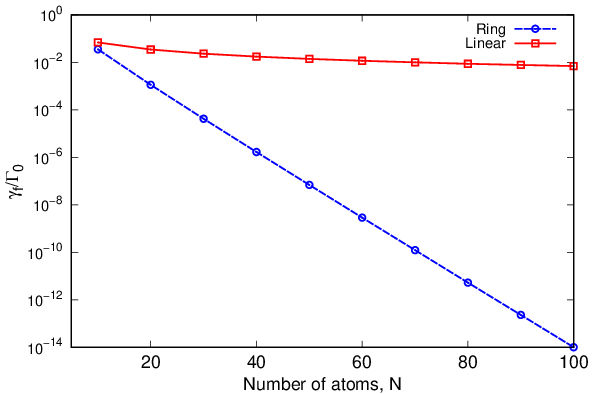}
    \caption{Decay rate of the TDS from an ordered atom array versus the number of atoms in the array. Red squares denote a linear array, and blue circles denote a ring-shaped array.}
    \label{fig:Ar_Ring}
\end{figure}

For the linear array that we have been discussing so far, most of the emission into free space occurs at the ends of the array. 
Reference \cite{ama2017} has shown an exponential improvement in subradiance when using a circular array configuration. 
The structure of the microring resonator naturally lends itself to the potential of having a ring of atoms trapped along the resonator.

Figure \ref{fig:Ar_Ring} shows the comparison between a linear array and a ring-shaped array trapped along the microring resonator. 
It depicts the decay rate into free space that is achievable using the TDS excitation through the waveguide as a function of the number of atoms in the array.
The comparison considers atoms trapped in arrays with $0.3 \lambda_0$ spacing, arranged in either a line or a ring above a hypothetical microring resonator that follows the array geometry, with uniform coupling at $C_1=0.05$. This comparison allows us to explore how the system scales with increasing atom number. In a practical setting, such a microring resonator could support a ring of about 400 atoms at the given spacing.
This is similar to Fig. 7 in Ref. \cite{ama2017}. We can almost reach a perfect subradiant state with the ring resonator setup and evanescent field traps.

A ring of atoms can form its own WGM, resulting in minimal loss of photons into free space. The interactions between this mode and the WGM of the resonator hold great potential for applications in quantum control and photon storage.

Although there is great promise with a ring-shaped array due to the exponential scaling, the effects of disorder described in Sec. \ref{sec:ArDisorder} will still limit the effectiveness of the system.

\section{Conclusion\label{sec:conclusion}}

We have studied the effects of collective dipole-dipole interactions in a system of atoms trapped in the vicinity of a nanophotonic microring resonator. We modeled and simulated the experimental setup in Ref. \cite{ExpPaper} to understand the dynamics and inform potential future research directions.
The atoms can interact with each other in two ways: the free space dipole-dipole interaction and the microring cavity interaction. Different properties and the mismatched wavenumber of the two lead to interesting dynamics. 

We studied the spectrum and decay rates of a cloud of atoms trapped near the resonator. We explored how the two different types of excitation, using a short pulse (for the timed-Dicke state) or a long pulse (for the steady state), affect the decay dynamics of the system. Although a suppression of the decay rate can be seen with the steady state, achieving selective radiance would be difficult in randomized atomic clouds.

Hence, we explored the implications of the proposed plan of trapping atoms in sub-wavelength arrays along the resonator. We show that this could afford the coherence necessary to achieve selective radiance. 
We studied the effects of the high-wavenumber excitation through the microring and explored potential impacts of disorder on subradiance in experimental implementations.

This versatile light-matter interaction platform holds great promise with its many degrees of control and tunable parameters. Implementing atom arrays will open up the possibility of "selective radiance." Expanding the study to three-level atoms opens possibilities to study applications such as photon storage.

The data plotted in the figures are openly available in Ref. \cite{datalink}.

\begin{acknowledgments}
The work of D.S. and F.R. was supported by the National Science Foundation under Award No. 2410890-PHY. The work of X.Z. and C.L.H. was supported by the AFOSR (Grant NO. FA9550-22-1-0031) and the ONR (Grant NO. N000142412184). This research was supported in part through computational resources provided by Information Technology at Purdue University, West Lafayette, Indiana.
\end{acknowledgments}

\appendix

\section{Adiabatic Elimination of the cavity\label{app:adiabaticElim}}
This appendix briefly describes the process of adiabatically eliminating the cavity photons from the Hilbert space due to the separation of timescales.

Beyond the cavity interaction Hamiltonian given in Eq. \eqref{eqn:H_I}, the cavity has two sources of loss. It can intrinsically lose a photon with the rate $\kappa_i$. Cavity photons can also hop to the bus waveguide with coupling rate $\kappa_e$.
This cavity decay is defined by the Lindblad operator,
\begin{equation}
    \mathcal{L}_{R}(\hat{\rho}) = (\kappa_i  + \kappa_e) [\hat{a} \rho \hat{a}^{\dagger} -\frac{1}{2} \hat{a}^{\dagger} \hat{a} \rho - \frac{1}{2} \rho \hat{a}^{\dagger} \hat{a}].
    \label{eqn:CavityLindblad}
\end{equation}

Since the cavity decay timescales are two orders of magnitude faster than the relevant atomic excitation timescales, they can be adiabatically eliminated.
By enforcing that the cavity reaches a steady state at every time-step of the internal dynamic evolution ($\langle\dot{\hat{a}}\rangle = 0$), we arrive at an expression for $\langle \hat{a} \rangle$ depending on the atomic excitations.

\begin{equation}
    \langle \hat{a} \rangle =  \frac{1}{\tilde{\kappa}}\bigg( \sum_j g_j e^{i\phi_j} \langle \hat{\sigma}_j^- \rangle + \eta \bigg).
    \label{eqn:ExpA}
\end{equation}

By replacing $\hat{a}$ in Eq. \eqref{eqn:H_I} and Eq. \eqref{eqn:CavityLindblad}, the cavity can be adiabatically eliminated, and the master equation can be re-written to arrive at Eqs. \eqref{eqn:Hddc}, \eqref{eqn:lindbladc} and \eqref{eqn:HL}.

\section{Orthogonality of the eigenmodes and final states\label{app:ortho}}

This appendix discusses the orthogonality relations in the combined Green's function matrix and demonstrates how they can be used to directly compute the final states. These unfamiliar orthogonality relations arise due to the different non-Hermitian properties of the relevant matrices.

The atom-cavity interaction matrix $\mathbf{G}_c$ is skew-symmetric because of the running-wave nature of the resonator. 
This results in purely imaginary eigenvalues, meaning none of the eigenmodes of the atom-cavity interaction has any energy shift. 
Any two eigenmodes $\vec{V}_{\alpha}, \vec{V}_{\beta}$ follows the familiar orthogonality relation: $\vec{V}_{\alpha}^{\dagger}\vec{V}_{\beta} = \delta_{\alpha\beta}$, where $\delta_{\alpha\beta}$ is the Kronecker delta.

On the other hand, the interaction matrix of the free space modes $\mathbf{G}_f$ is complex-symmetric. This means that $G^{f}_{ij} = G^{f}_{ji}$. This results in the eigenmodes having a different orthogonality relation: $\vec{V}_{\alpha}^T\vec{V}_{\beta} = \delta_{\alpha\beta}$, where $T$ denotes a transpose without conjugation. 

The combined matrix will be neither Hermitian nor Complex-symmetric, resulting in different left $(\vec{L})$ and right eigenvectors $(\vec{R})$ and a different orthogonality relation. 
\begin{equation}
 \begin{split}
     \mathbf{M} \vec{R}_{\alpha} = \lambda_\alpha \vec{R}_{\alpha} \\
     \vec{L}_\alpha^T \mathbf{M} = \lambda_\alpha \vec{L}_\alpha^T,
 \end{split}   
\end{equation}
where $\lambda_\alpha$ denotes the eigenvalue of eigenmode $\alpha$ of the full interaction matrix. 
The left and right eigenvectors $\vec{L}_\alpha$ and $\vec{R}_\alpha$ can then be normalized to follow
\begin{equation}
    \vec{L}_\alpha^{T} \vec{R}_{\beta}= \delta_{\alpha\beta}.
\end{equation}

Using this orthogonality relation, we can decompose the driving laser term in Eq. \eqref{eqn:MFTEqn} into components that each drive a particular eigenmode $\alpha$
\begin{equation}
    \tilde{\Omega}_\alpha = \vec{L}_\alpha^{T} \vec{\Omega}.
\end{equation}

Since each eigenmode evolves individually with its characteristic decay rate, the equation of motion of each eigenmode can be solved separately. 
The contribution from each eigenmode $w_\alpha$ can be calculated for the SS in the low intensity,
\begin{equation}
    w_\alpha^\mathrm{SS} = - \frac{\tilde{\Omega}_\alpha}{2(\lambda_\alpha -  \Delta_A)},
\end{equation}
and for the TDS
\begin{equation}
    w_\alpha^\mathrm{TDS} \propto \tilde{\Omega}_\alpha,
\end{equation}
with the proportionality depending on the pulse area. These can be used with Eq. \eqref{eqn:sigmavec} to calculate the final state and the composition with respect to the eigenmodes.

\bibliography{ref.bib}
 
\end{document}